\documentclass[preprint,12pt]{elsarticle}
\usepackage{amsmath,amssymb,amsfonts}
\usepackage{algorithmic}
\usepackage{algorithm}
\usepackage{array}
\usepackage{textcomp}
\usepackage{stfloats}
\usepackage{url}
\usepackage{verbatim}
\usepackage{graphicx, caption, subcaption}
\usepackage{xcolor}
\usepackage{color,soul}
\usepackage{pifont}
\usepackage{multirow}
\usepackage{multicol}
\usepackage{booktabs}
\usepackage{amssymb}
\usepackage{amsmath}
\usepackage{comment}
\usepackage{multirow}
\usepackage{multicol}
\usepackage{booktabs}

\journal{Nano Communication Networks}

\begin{document}

\begin{frontmatter}

\title{TRMAC: A Time-Reversal-based MAC Protocol for Wireless Networks within Computing Packages}

%% Authors and their footnotes
\author{Ama Bandara\corref{cor1}}
\author{Abhijit Das}
\author{F\'atima Rodr\'iguez-Gal\'an}
\author{Eduard Alarc\'on}
\author{Sergi Abadal}

%% Corresponding author note
\cortext[cor1]{Corresponding author: Ama Bandara (\texttt{ama.peramuna@upc.edu})}

%% Acknowledgement footnote
\fntext[fn1]{Authors gratefully acknowledge funding from the European Commission via projects with GA 101042080 (WINC), 101099697 (QUADRATURE) and 101189474 (EWiC).}
%% Affiliations
\address{Universitat Politècnica de Catalunya, Barcelona, Spain
}

%% Abstract
\begin{abstract}
As chiplet-based integration and many-core architectures become the norm in computing, on-chip wireless communication has emerged as a compelling alternative to traditional interconnects. However, scalable Medium Access Control (MAC) remains a fundamental challenge, particularly under dense traffic and limited spectral resources. This paper presents TRMAC, a novel cross-layer MAC protocol that exploits the spatial focusing capability of Time Reversal (TR) to enable multiple parallel transmissions over a shared frequency channel. By leveraging the quasi-deterministic nature of on-chip wireless channels, TRMAC pre-characterizes channel impulse responses to coordinate access using energy-based thresholds, eliminating the need for orthogonal resource allocation or centralized arbitration. Through detailed physical-layer simulation and system-level evaluation on diverse traffic, TRMAC demonstrates comparable or superior performance to existing multi-channel MAC protocols, achieving low latency, high throughput, and strong scalability across hundreds of cores. Moreover, we prove that TRMAC can be utilized for parallel transmissions with a single frequency channel with a similar throughput and latency as in using multiple frequency bands, omitting the need for complex transceivers.
\end{abstract}

%% Keywords

\end{frontmatter}

%% Main Text
% Introduction
\section{Introduction}~\label{sec:intro}

Recent advances in computer architecture have led to the integration of many processing cores within a single computing package, and more recently, the disaggregation of functionality across \emph{chiplets} interconnected via high-density packaging \cite{salahuddin2018era}. To fully harness the computational capabilities of such architectures, whether monolithic or chiplet-based, efficient parallel processing has become indispensable to enable the concurrent execution of tasks and efficient handling of large-scale data. However, this parallelism introduces significant challenges in the communication across cores, within or across chiplets, that is used to maintain synchronization and ensuring coherent data sharing within the system \cite{das2024chip, Karkar2016, Shamim2017}. Consequently, precise and efficient communication mechanisms are critical for coordinating tasks and maximizing system performance \cite{bertozzi2015, nabavinejad2020overview, Karkar2016}.

Wireless Networks-on-Chip (WNoCs) have recently emerged as a promising complement to traditional wired NoC architectures, offering a path to alleviate the growing challenges of latency and power dissipation in massive System-on-Chip (SoC) or System-in-Package (SiP) designs \cite{bertozzi2015, Sujay2012}. In these systems, wireless communication can be facilitated through antennas that are integrated on each core or chiplet and that establish wireless links within the package. Compared to conventional wired interconnects, wireless links offer substantial advantages, including low latency via single-hop data transfers, inherent broadcasting capabilities, and dynamic reconfigurability \cite{Sujay2012, timoneda2020engineer}. However, to fully exploit the benefits of wireless interconnections in high-performance environments,
future WNoCs designs would target data rates up to 100\,Gb/s, latencies below 10\,ns, energy consumption around 1\,pJ/bit, and Bit Error Rates (BERs) below $10^{-10}$ \cite{abadal2018medium}.

While the use of wireless communication within chip packages offers compelling benefits, it also presents unique challenges. The confined and reverberant nature of the chip environment leads to significant delay spreads due to many reflections, resulting in high Inter-Symbol Interference (ISI) that degrades communication at high data rates. Furthermore, since signals spread across the entire environment, severe Co-Channel Interference (CCI) can be expected for transmissions within the same time-frequency channel. 

\begin{figure}[t]
\centering
\includegraphics[width=\textwidth]{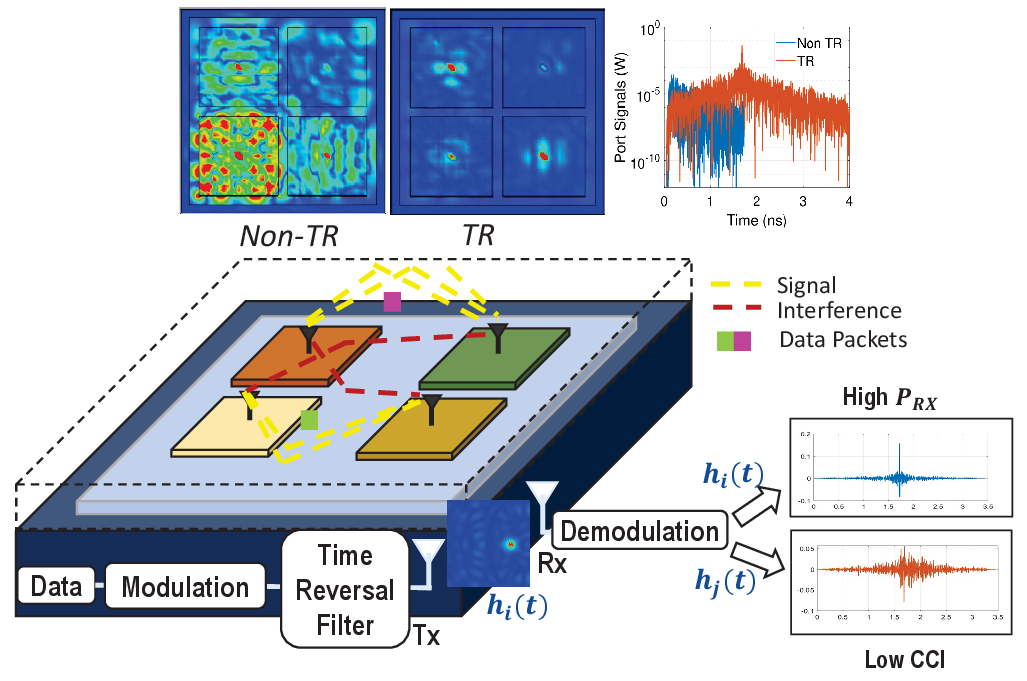}
\caption{Illustration of the scenario considered in this paper. Wireless Network-on-Chip (WNoC) in a multi-chip package with multiple parallel communication links. Data packets are transmitted using time reversal, which may suppress co-channel interference at neighboring nodes. Spatial channels need to be coordinated via a link-level protocol, for which we propose TRMAC in this work.}
\label{fig:wnoc}
\end{figure}

Fortunately, the quasi-deterministic nature of the on-chip wireless channel, enabled by fixed package dimensions and known material properties, allows to mitigate the ISI and CCI impairments via a technique called Time Reversal (TR). TR is a signal processing technique that employs the time-reversed version of the channel response between a transmitter-receiver pair as a spatial matched filter. As a result, TR concentrates the radiated energy at the vicinity of the intended receiver and around a given time instant, as shown in Figure~\ref{fig:wnoc}, thereby significantly reducing ISI and enhancing signal fidelity~\cite{bandara2023exploration, rodriguez2023, trcommag}. Additionally, TR minimizes CCI by exploiting the orthogonality of spatial channels between different transmitter-receiver pairs. This allows, in principle, multiple parallel transmissions to occur simultaneously in both time and frequency domains. 

The static nature of the wireless channels in this scenario makes TR a very attractive technique for improving communication robustness, since the channel response needs to be measured just once per each transmitter-receiver pair for the creation of the TR filter bank. Because the channels' responses do not change over time, the filter bank can be reused throughout the lifetime of the computer to mitigate ISI and CCI. \textbf{However, the use of TR filters needs to be carefully managed at runtime since, 
despite enabling spatial multiplexing, TR does not inherently prevent transmission collisions.} Hence, to ensure robust and efficient data exchange, it is crucial to adopt Medium Access Control (MAC) that avoid or remedy collisions.

Existing MAC protocols for WNoC typically rely on resource-sharing schemes such as frequency-division, time-division, or code-division multiple access to mitigate interference in multi-link environments~\cite{Vijayakumaran2014, MatolakF2012}. Some approaches adopt hierarchical hybrid wired-wireless architectures, but tend to emphasize overall system design while overlooking the effects of wireless channel impairments~\cite{zhao2008, Ganguly2011, DiTomaso2015, yu2014architecture, Shamim2017}. More recent protocols apply combinations of multiple access techniques and propose intelligent techniques to minimize collisions \cite{Shamim2017, mansoor2015reconfigurable, franques2021fuzzy, mestres2016mac, Vijayakumaran2014, MatolakF2012, DiTomaso2015, Ollé2023, jog2021one, Rout2023}, as summarized in Section \ref{sec:relatedwork} and more particularly in Table~\ref{tab:relatedwork}.

However, despite significant progress in WNoC, the existing MAC protocols do not account for the unique physical layer characteristics of the chip environment. In this context, TR represents a huge opportunity since it enables spatial multiplexing in such a reverberant environment without the need of antenna arrays; yet still, none of the existing MAC protocols supports it. In fact, the literature of MAC protocols supporting time reversal is extremely scarce in any environment. To the best of our knowledge, only the work of \mbox{\cite{Zhao2019}} explores a TR-based MAC for underwater acoustic networks. However, the underwater channel is highly time-variant and the communication is limited by the propagation latency and the use of acoustic transducers (as opposed to the static, short-range and high-frequency nature of WNoC).

In this paper, we propose TRMAC, a cross-layer adaptive MAC protocol that is specifically designed to leverage the unique characteristics of TR applied to on-chip wireless environments. TRMAC exploits, for the first time, the spatial multiplexing capabilities of TR to increase the throughput of wireless on-chip networks without requiring extra spectral resources. To that end, our protocol addresses the main challenge of TR: the deafness problem associated with the focusing ability of TR, which makes nodes unaware of other nodes' transmissions and that can lead to collisions. Collision management is performed in a fully distributed manner, using tone signaling, exponential backoff, and a unique mid-transmission acknowledgment mechanism that also exploits TR. In so doing, TRMAC is able to scale the aggregate bandwidth with the number of spatial channels, which could be combined with orthogonal multiplexing methods to further improve performance in massively multi-core and multi-chiplet systems.

In our previous work \cite{Bandara2024}, we presented the fundamental mechanisms behind TRMAC and an initial performance evaluation for different numbers of spatial channels, frequency channels, and number of antennas. In this paper, we significantly extend the work by providing a deeper description of the different phases of the protocol, potential collision situations, physical layer aspects, together with a more comprehensive evaluation that now includes traffic patterns of variable spatial hotspotness and temporal burstiness, and an extended discussion of the protocol overheads. In summary, the main contributions of this work are summarized as follows:

\begin{itemize}
    \item We propose TRMAC, a cross-layer MAC protocol that leverages TR filtering to enable parallel, low-interference wireless transmissions in chip-scale environments.
    \item We evaluate TRMAC under diverse traffic scenarios, including even, uneven, and bursty distributions, across 64 to 256 cores, demonstrating its scalability and robustness in dense multi-core and multi-chiplet systems.
    \item We discuss the strengths and limitations of the proposed approach and potential ways for future implementation of TRMAC in real systems.
\end{itemize}

The rest of the paper is organized as follows. Section~\ref{sec:background} provides the background for wireless on-chip communication and TR-based design. Section~\ref{sec:protocol} details the proposed TRMAC protocol and Section~\ref{sec:methodology} discuss its physical-layer implementation. Section~\ref{sec:results} presents the performance evaluation of TRMAC against existing approaches. Section~\ref{sec:relatedwork} reviews relevant MAC protocols in the context of WNoC architectures. Section~\ref{sec:discussion} provides a discussion of the overall results as well as of the limitations of the study. Finally, Section~\ref{sec:conclusion} summarizes the contributions and outlines directions for future work.

\newpage

% Background and Motivation
\section{Background}~\label{sec:background}
\subsection{Wireless Network-on-Chip (WNoC)}~\label{wnoc}
WNoC, augment traditional wired NoC infrastructures by overlaying wireless links, offering benefits such as single-hop communication, efficient broadcasting, and reconfigurability. These features are particularly valuable in large-scale systems, where dense routing and increased hop counts can lead to significant latency and power overheads. Moreover, with the rise of chiplet-based integration, where multiple heterogeneous dies are interconnected within the same package, the inter-chip  communication links are less efficient than the intra-chip links, 
for which wireless links offer a low-latency and scalable alternative to traditional interconnect fabrics. The monolithic or chiplet-based nature of on-package systems enables controlled ElectroMagnetic (EM) propagation through well-characterized materials and structures, making it possible to tailor channel behavior using pre-defined CIRs. As a result, WNoC architectures have seen the integration of various co-designed antennas and transceivers optimized for both core-to-core and chiplet-to-chiplet wireless communication~\cite{bellanca2017integrated, Rayess2017, timoneda2018channel, gutierrez2009chip, yeh2012design, hwangbo2019integrated}.

To meet high-performance requirements, such as ultra-low latency and high throughput, WNoCs must efficiently utilize available bandwidth. However, modulation schemes are typically constrained by power and area, limiting practical deployments to simple schemes like On-Off Keying (OOK) \cite{yu2014architecture}. Consequently, careful MAC is necessary to coordinate transmissions across shared resources and reduce contention. Several MAC protocols adapted from traditional wireless networking, such as token passing, CSMA, and TDMA, have been explored for WNoC environments as mentioned in Section \ref{sec:relatedwork} \cite{duraisamy2015enhancing, mansoor2015reconfigurable, Ollé2023, franques2021fuzzy, DiTomaso2015, mestres2016mac}. 
While these approaches effectively manage channel access and collision avoidance, they rely on orthogonal resource allocation and often assume full awareness of ongoing transmissions. They fall short in exploiting the spatial reuse potential of the on-chip environment and generally overlook the deafness problem that arises in non-orthogonal, spatially diverse settings.

\begin{figure}[t]
\centering
\begin{subfigure}[t]{0.4\textwidth}
\includegraphics[width=\textwidth]{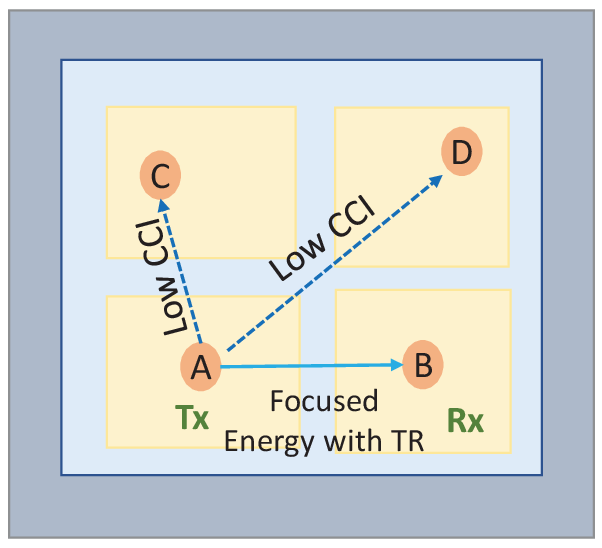}
\caption{}
\label{fig:trcci1}
\end{subfigure}
\hfill
\begin{subfigure}[t]{0.55\textwidth}
\includegraphics[width=\textwidth]{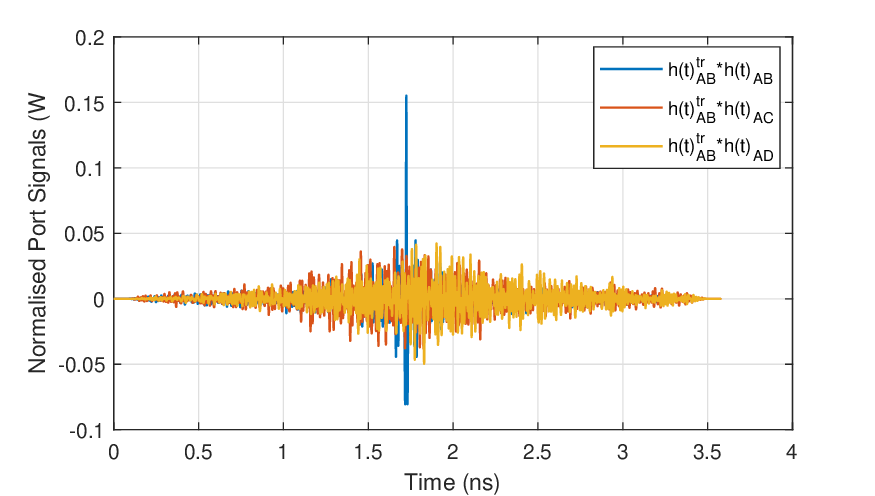}
\caption{}
\label{fig:trcci2}
\end{subfigure}
\vspace{0.2cm}
\caption{(a) TR communication between node A and node B with reduced CCI at nodes C and D. (b) Spatial and temporal focusing at the intended receiver (node B), with minimal interference at others.}
\label{fig:trcci}
\end{figure}

\subsection{Time Reversal (TR) for On-Chip Communications}~\label{TR}
TR is a physical-layer technique that focuses EM energy at a specific receiver by transmitting a signal filtered through the time-reversed version of the CIR. This process effectively leverages the multipath richness of reverberant environments to achieve spatial and temporal energy focusing. Given that both monolithic and chiplet-based packages exhibit quasi-deterministic layouts and high reverberation, TR is a highly promising technique to mitigate multipath-induced impairments.

TR offers two key benefits for on-chip communication: (1) suppression of ISI by temporal focusing, and (2) mitigation of CCI via spatial discrimination. Each transmitter–receiver pair has a unique CIR, resulting in low cross-correlation between links. As shown in Figure~\ref{fig:trcci}, a signal from node A pre-filtered with the time-reversed CIR to node B will be sharply focused at B, while nearby nodes (e.g., C and D) receive minimal interference. This spatial orthogonality enables multiple simultaneous transmissions over the same frequency band without the need for traditional orthogonalization.

\subsection{Medium Access Control (MAC) Design in WNoC}

In WNoCs, MAC protocols play a critical role in balancing latency, throughput, and energy efficiency. While traditional MAC strategies rely on orthogonal access to prevent interference, the spatial focusing offered by TR opens up new opportunities for non-orthogonal, spatially multiplexed communication. However, this also introduces new challenges such as managing interference from partially overlapping spatial channels and addressing the deafness problem, where nodes are unaware of ongoing transmissions due to focused, rather than broadcast, communication.

The design of effective MAC protocols for WNoC must account for the deterministic and static nature of the environment, enabling pre-characterized CIRs to guide communication decisions. By combining this channel knowledge with lightweight control mechanisms, it becomes feasible to manage contention across shared frequency and time resources, even in high-density multi-core and multi-chiplet systems. Additionally, the MAC must adapt to dynamic and uneven traffic patterns, which are typical in modern heterogeneous computing workloads. Therefore, simplicity, adaptability, and scalability are key requirements for practical on-chip MAC protocol designs.

% Overview of Protocol
\section{The TRMAC Protocol}~\label{sec:protocol}
Efficient data sharing and synchronization are critical in multi-core and multi-chiplet systems. In massive SiP architectures, parallel transmissions can reduce communication latency, but limited bandwidth must be carefully shared across concurrent links, necessitating orthogonal resource allocation (e.g., frequency or time division). Existing MAC protocols for WNoCs typically rely on FDMA or TDMA to multiplex access, which either increases latency or divides limited bandwidth among competing transmissions. These approaches do not take advantage of the spatial diversity available in on-chip wireless environments. With the introduction of TR filtering, spatial multiplexing becomes feasible. By exploiting distinct CIRs among node pairs, multiple concurrent transmissions can occur on the same frequency and time without orthogonal separation, significantly increasing aggregate bandwidth and reducing latency.

Motivated by this opportunity, we introduce TRMAC, a cross-layer MAC protocol that leverages pre-characterized CIRs and the deterministic nature of on-chip channels to support parallel transmissions through spatial reuse. TRMAC establishes per-node energy thresholds and uses preamble-based signaling to manage collisions and mitigate deafness. Because it is grounded in physical-layer channel knowledge, TRMAC can be combined with FDMA or TDMA to further increase spatial reuse, offering a practical and scalable solution for high-density multi-core and multi-chiplet systems.

To explain the protocol, we first provide an overview of its phases in Section \ref{sec:structure}, then clarify the main design decisions and assumptions in Section \ref{sec:decisions}, and finally describe the collision management mechanisms in Section \ref{sec:manage}.

\begin{figure}[t]
\centering
\includegraphics[width=1\columnwidth]{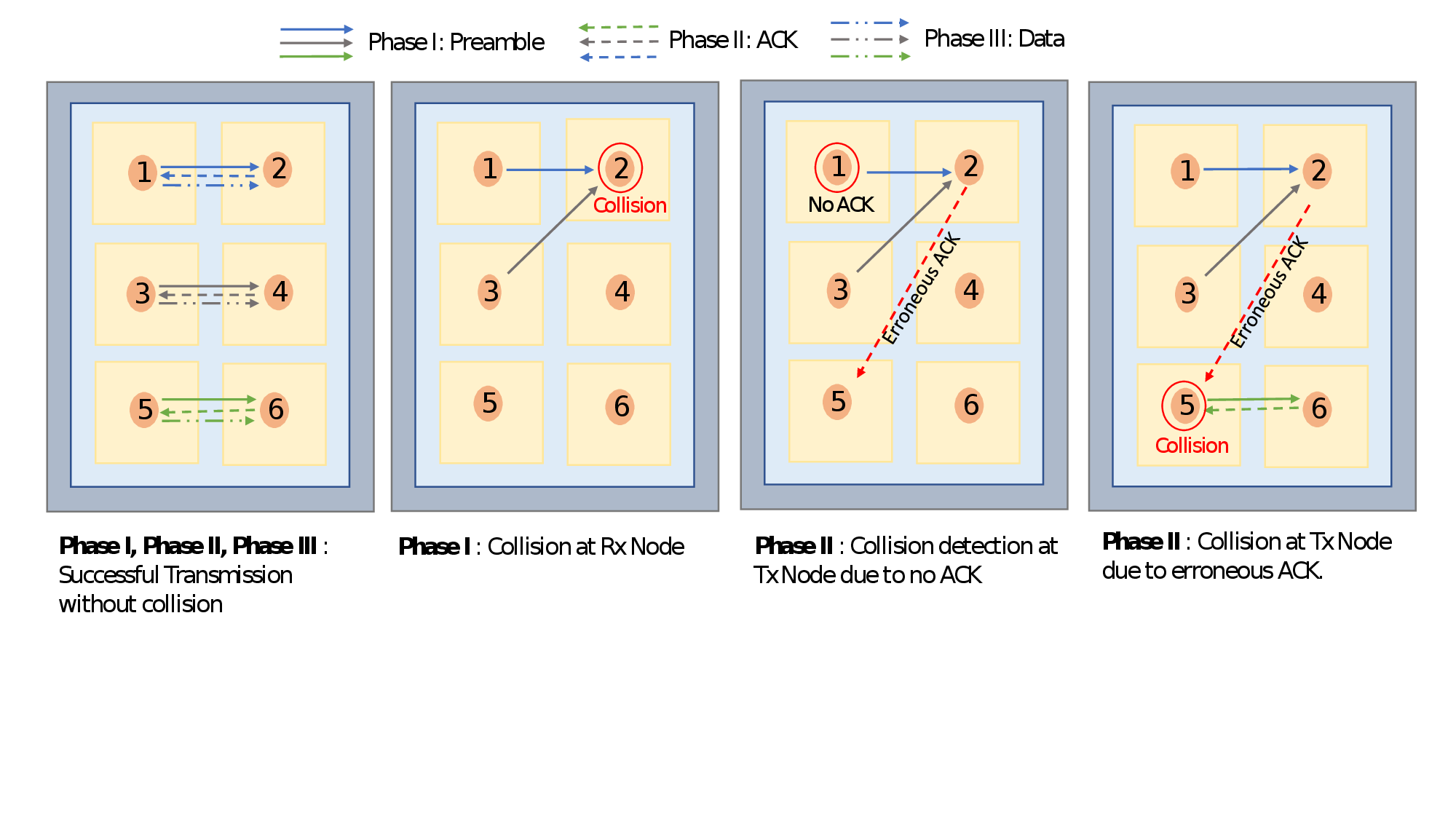}
%\vspace{-0.2cm}
\caption{Illustration of TRMAC: (Left) successful parallel transmissions. (Middle left) preamble collisions at node 2. (Middle right) missed ACKs from Phase II. (Right) erroneous ACK affects parallel transmission.}
\label{fig:collision}
\end{figure}

\begin{figure}[t]
\centering
\includegraphics[width=0.7\columnwidth]{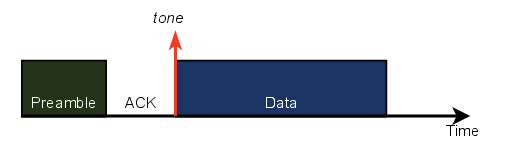}
%\vspace{-0.3cm}
\caption{Timing diagram of a single transmission with TRMAC.}
\label{fig:timing}
\end{figure}

\subsection{Overview of the Protocol}
\label{sec:structure}
The TRMAC protocol operates in three sequential phases: preamble transmission, ACK detection, and data transfer. If a node is ready to transmit and the channel is idle, we first transmit the address of the Transmitter (Tx) to the Receiver (Rx) as the preamble, using the appropriate TR filter. The Rx receives and decodes the address from the preamble, and then transmits the Acknowledgment (ACK) to the original sender using the TR filter pointed by the source address. If there is no collision, the Rx will use the correct TR filter and hence the Tx will receive the ACK. In this case, the Tx will continue with the rest of the data transmission together with a tone signal to avoid further collisions at this stage. If there was a collision in the first stage, the Rx will decode an incorrect address and the protocol will have to start over, as further explained in subsequent sections.

\begin{figure}[htbp]
\centering
\includegraphics[width=1\columnwidth]{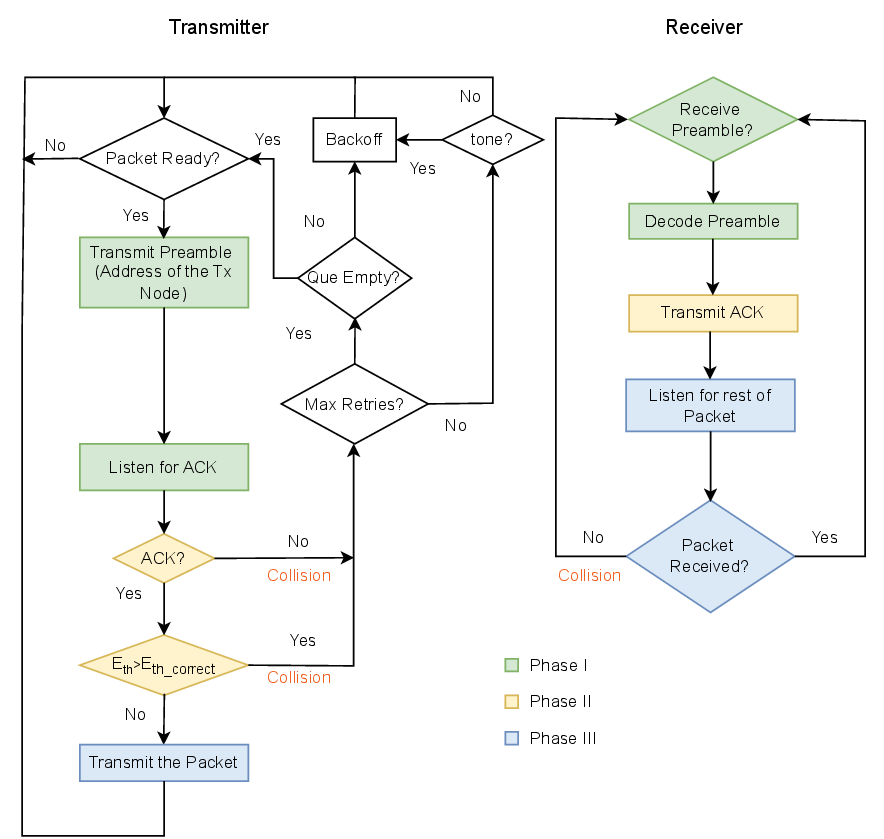}
%\vspace{-0.3cm}
\caption{Flowchart of the TRMAC protocol.}
\label{fig:algo}
\end{figure}

These three stages are illustrated in Figures~\ref{fig:collision} and~\ref{fig:timing}, while the detailed working flowchart is presented in Figure~\ref{fig:algo}. In all cases, it is considered that all nodes have a TR filter bank that is filled at the system startup time using calibration measurements. Since the system is static, the filter bank is valid for the entire system lifetime.

\subsubsection{Phase I: Preamble Transmission} \label{preamble}
\begin{figure}[t]
\centering
\includegraphics[width=1\columnwidth]{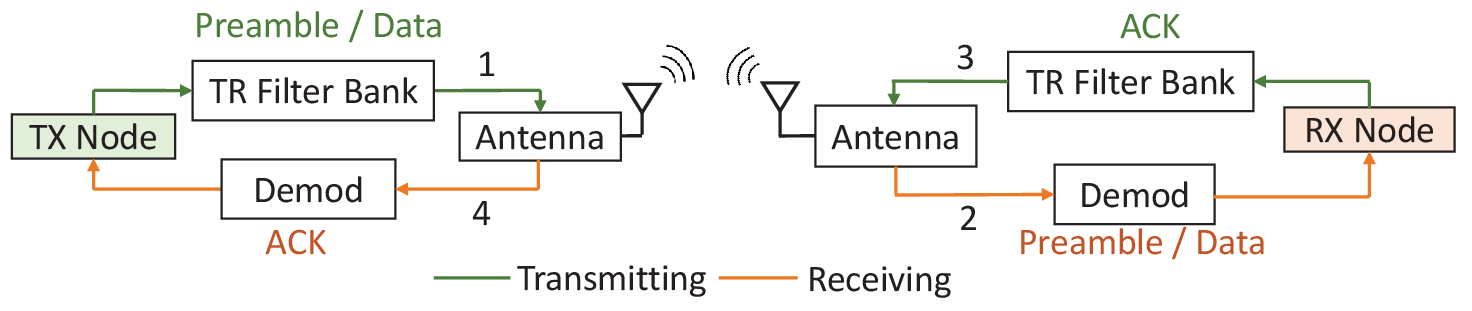}
%\vspace{-0.3cm}
\caption{Block diagram illustrating the transmitting and reception process in TRMAC. Tx is on the left and Rx is on the right. (1) The Tx node first transmits the time reversed preamble through the TR filter bank. (2) The Rx node demodulates the preamble and decodes the source address. (3) The Rx node transmits the ACK going through the TR filter bank based on the address demodulated in the preamble. (4) The Tx node receives the ACK, after which the Tx can transmit the rest of the data.}
\label{fig:txrx}
\end{figure}

When a packet is ready and the channel is idle, the Tx node initiates communication by sending a preamble, filtered through the time-reversed CIR of the intended Rx node as illustrated in inset 1 of Figure \ref{fig:txrx}. The preamble contains the Tx node’s address. Then, the protocol proceeds to the next phase (Section \ref{phase2}).

\subsubsection{Phase II: Acknowledgment Detection} \label{phase2}
Upon receiving the preamble, the Rx decodes the address of the Tx, selects the corresponding CIR on its filter bank, and sends back a TR-based ACK. Insets 2 and 3 of Figure \ref{fig:txrx} depict the reception of the preamble and the transmission of the ACK with TR. Due to spatial focusing, the ACK should reach only the intended Tx node, which is waiting for the ACK after the preamble transmission. Two situations may occur:

\begin{itemize}
    \item \textbf{No Collision:} If there is no collision, the ACK reaches the intended Tx node, which then proceeds to the next phase of the protocol (Section \ref{phase3}). This is illustrated in the inset 4 of Figure \ref{fig:txrx}. 
    \item \textbf{Collision:} If two transmitters try to connect to the same Rx, a collision may occur. The Rx will try to decode the source address from a signal that contains the overlap of two transmissions, and probably lead to an incorrect address. Therefore, the ACK will most likely reach nodes not originally involved in the communication. Another type of collision can happen if the intended Rx is busy trying to transmit elsewhere and misses the preamble. In both collision cases, the Tx does not receive any ACK; the Tx concludes that a collision may have occurred and halts further transmission, looping back to the first phase of the protocol after an exponential backoff phase.  
\end{itemize}

\subsubsection{Phase III: Data Transmission}
\label{phase3}
If the Tx receives an ACK that matches the instant when the preamble transmission was made (inset 4 of Figure \ref{fig:txrx}), it proceeds to transmit the rest of the data packet to the Rx using the appropriate TR filter. Also, during this phase, the Tx activates its tone (see Figure \ref{fig:timing} and Section \ref{sec:manage}) to avoid any other nodes starting a transmission until the transmission is complete. This ensures that collisions cannot occur in this phase.

\subsection{Design Decisions}
\label{sec:decisions}
A comprehensive illustration of the operation of TRMAC from the perspective of the transmitter and receiver is shown in Figure~\ref{fig:algo}. Next, we provide some additional details on the design decisions made.

\begin{itemize}
    \item \textbf{Transmitter:} If we wish to extend TRMAC to multiple frequency channels, one can initially assign $f_n$ frequency channels  are uniformly distributed among nodes, with each channel covering a sector of $N_{\text{cores}}/f_n$. Transmissions are assumed to be globally synchronized. Each Tx node starts communication from Phase I and proceeds sequentially unless a collision is detected. To detect ACKs and possible collisions, each Tx compares the received signal energy to a predefined threshold $E_{\text{th,correct}}$. If a Tx aims to communicate with multiple nodes simultaneously through symmetric spatial channels, this threshold is adjusted using pre-calculated values to reflect the expected composite energy.
    \item \textbf{Receiver:} Upon receiving a preamble, the Rx uses threshold-based detection to identify the Tx. If multiple transmissions arrive simultaneously at the same Rx, the energy may superimpose, leading to incorrect preamble decoding and misidentification of the Tx. Thus the Rx may select the wrong CIR, causing a failed transmission and contributing to overall collision rate. Additionally, if the intended Rx initiates its own transmission at the same time and frequency, self-interference prevents it from receiving the original preamble or sending an ACK, again contributing to the deafness problem.
\end{itemize}

\subsection{Collision Management} \label{collision}
\label{sec:manage}
In previous sections, we mention that a transmitter can start a transmission if a packet is ready and the channel is idle. For this to occur, a transmitter should not detect a busy tone or be on a backoff period. These are two collision management mechanisms that are further explained next.

\begin{itemize}
    \item  \textbf{Tone Signaling:} If there is no collision in the first two phases, we move to the phase of data transmission. In this phase, the transmission of the payload occurs, which is generally longer than the transmission of the preamble. However, other nodes may not be aware that this transmission is occurring, because TR makes the preamble and ACK exchange to only be seen by the intended Tx and Rx thanks to its spatiotemporal focusing ability. To avoid other transmitters to attempt a preamble transmission at this point, a continuous tone is broadcast over a separate, dedicated, low-bandwidth channel. This communicates that the Phase III of the protocol is on-going in one or more transmissions. If no tone is detected, the medium is assumed to be free.
    \item \textbf{Exponential Backoff:} To further clarify the protocol operation when a collision occurs, let us consider the transmission scenario illustrated in Figure \ref{fig:collision}, where in Phase I, node 1 and node 3 simultaneously transmit their time-reversed preamble data to node 2. Here, with the collision occurring in node 2, the respective address of the Tx node will be erroneously detected in the physical layer. Therefore, as indicated in Figure \ref{fig:collision}, following the previous collision, an erroneous ACK will be sent to a neighboring node in the system, while the Tx node will not receive any ACK. In this scenario, as the Tx node does not receive any ACK, it identifies the possible collision and follows a process of exponential backoff, i.e. enters a wait period that grows exponentially with each consecutive collision, reducing the probability of repeated interference during retries.
\end{itemize}

\begin{figure}[t]
\centering
\includegraphics[width=0.7\textwidth]{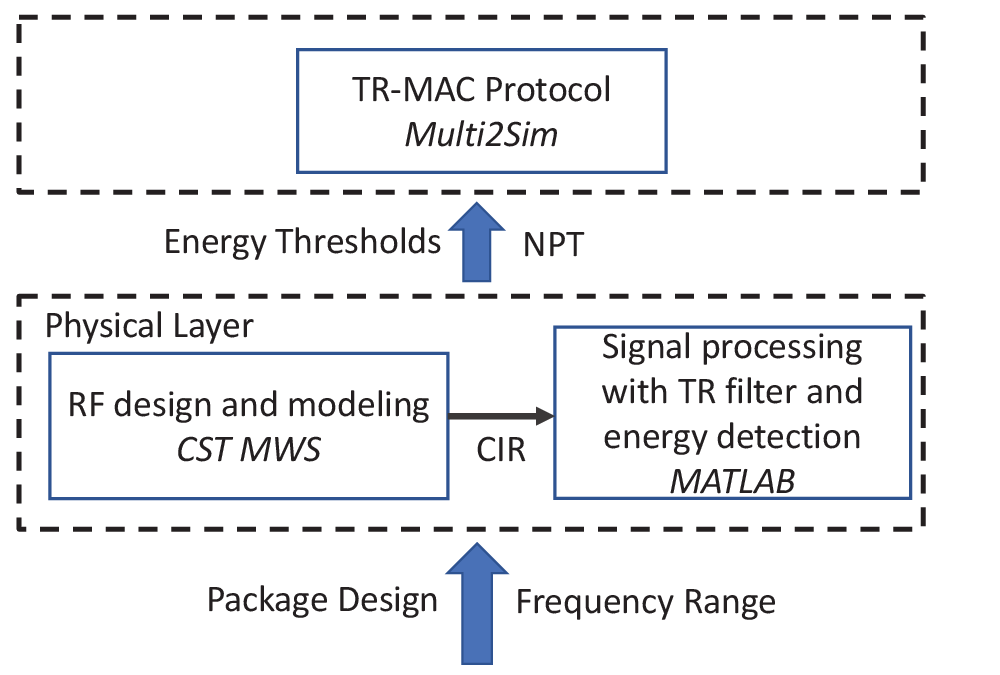}
\caption{TRMAC framework: CIR pre-characterization enables energy-focused communication and decentralized collision management via preamble signaling and energy threshold detection.}
\label{fig:trframework}
\end{figure}

%methodology
\section{Methodology}~\label{sec:methodology}
The protocol leverages the unique characteristics of the on-chip wireless channel by starting from the physical layer. Therefore, the methodology of this work considers channel and physical layer simulations prior to the link-level simulations, as illustrated in Figure~\ref{fig:trframework}. 

First, the chip package is simulated using CST Microwave Studio \cite{cst} to extract the channel responses. These channel responses are pre-characterized and considered static, assuming a fixed chip layout and material composition. On an actual implementation of the system, only a single-time calibration or measurement phase would be needed to obtain the channel for all antenna pairs. The channel is time-invariant, therefore those channel measurements are used during the lifetime of the device and there is no need to estimate the channel with repetitive probing.

Second, MATLAB \cite{MATLAB_R2024b} simulations are used to implement a basic Amplitude Shift Keying (ASK) modulation and to calculate energy thresholds for bit detection with low BER. We then estimate the Number of Parallel Transmissions (NPT) that can coexist without exceeding acceptable BER limits, based on cross-interference among simultaneous links. Importantly, as discussed in Section~\ref{wnoc}, both the energy thresholds and NPT are fixed once the chip is designed and remain stable due to the static nature of the in-package channels. Finally, these parameters are incorporated into Multi2Sim \cite{ubal2012multi2sim} to evaluate protocol feasibility under practical operating conditions.

\begin{figure}[t]
\centering
\includegraphics[width=0.8\columnwidth]{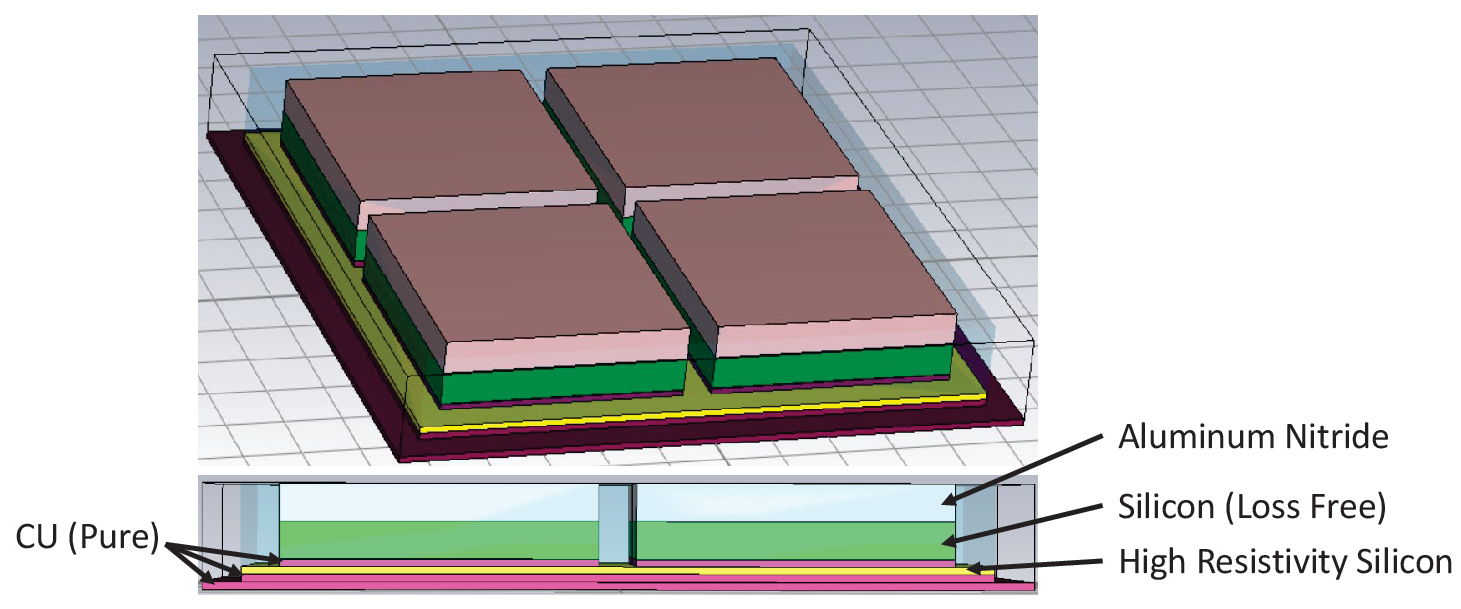}
%\vspace{-0.3cm}
\caption{Cross section and top view of a package with four chiplets of each 5×5 mm\textsuperscript{2} each, modeled in CST Microwave Studio. Vertical monopole antennas are placed through the silicon layer at a horizontal distance in the range of 1.29--4.38$\lambda$ of each other.}
\label{fig:cross_sec}
\end{figure}

\begin{figure}[t]
\centering
\includegraphics[width=0.5\columnwidth]{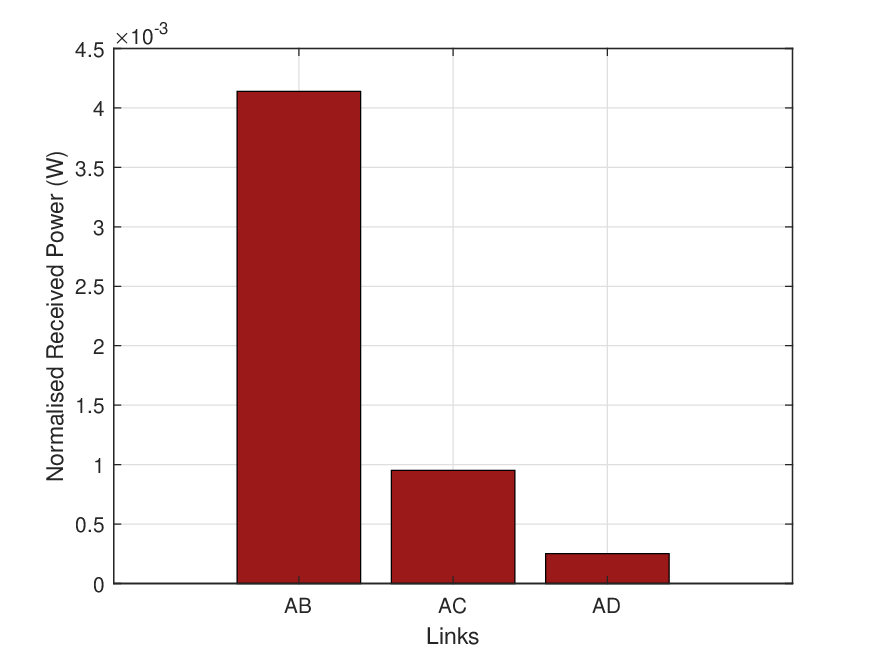}
%\vspace{-0.3cm}
\caption{Total received power from A to the intended receiver B and non-intended receivers C and D}
\label{fig:interference}
\end{figure}

\subsection{Physical Layer Simulation}
To model the on-chip wireless environment, a four-chiplet interposer package was simulated using CST Microwave Studio~\cite{cst}, with each core equipped with a single antenna operating at a center frequency of 140\,GHz. The details of the flip-chip layers follows the setup described in~\cite{rodriguez2022towards}. The interposer is created by stacking four flip-chip structures on top of silicon interposer and copper layers. Vertical monopole antennas are placed on each chiplet in symmetrical positions such that the cylindrical metallic structure is passed through the silicon substrate and excited from the bottom with metal layer acting as the ground plane. The choice of a vertical monopole is motivated by its increased co-planar coupling. The length of the monopole is obtained by considering the antenna center frequency and the thickness and relative permittivity of the silicon substrate \mbox{~\cite{rodriguez2022towards}}. The model we built consists of four 5×5 $mm^{2}$ chiplets as shown in \mbox{Figure \ref{fig:cross_sec}}. The distances between the  antennas are in the range of $1.29\lambda-4.38\lambda.$

After accurate antenna modeling, the CIR is obtained by sequential impulse excitations followed by the antennas, where we can extract the information of each link for post-processing. Theoretically, TR is compatible with any antenna design, as through the channel characterization entire channel response is obtained, including the antennas themselves, to calculate the filter's response. In MATLAB, serialized data is modulated using ASK, passed through a TR filter (complex conjugate of the CIR), and transmitted through the wireless channel. The receiver captures the signal and decodes the data by comparing received energy to pre-characterized energy thresholds~\cite{bandara2023exploration, rodriguez2023}.

With this model we simulate and analyze the cross-interference per each link by following \mbox{Figure \ref{fig:trcci1}}. The results in \mbox{Figure \ref{fig:interference}} show us the received energy per each node in a single transmission with TR. When the signal is focused to the intended receiver as in link $AB$, the received power is high due to focused energy. However, a residual CCI is obtained as shown in links $AC$ and $AD$ and due to the cross-correlation between the links $AB-AC$ and $AB-AD$. Yet, when multiple transmissions are utilized, the impact of CCI on each node is accumulated and consequently the BER is increased. However, with pre-characterized channel CIR we can determine the number of parallel transmissions (i.e. NPT) that can be reliably supported while maintaining BER below an acceptable threshold.

\subsection{Link Layer Simulation}
To assess the performance of TRMAC, several link-layer aspects need to be clarified and incorporated in Multi2sim. For instance, even though wireless data transfer in TRMAC can be performed over multiple frequency channels, this work focuses on a single frequency channel. The timing of all transmissions is considered to be aligned to a global system clock running at 1 GHz. Therefore, Multi2sim simulations are slotted at a nanosecond granularity. In this context, we consider the overall transmission to take $N+2$ clock cycles: one for the preamble transmission, one for the acknowledgment, and $N$ cycles for the transmission of the data.

In our particular set of simulations, each packet consists of 80 bits, which is consistent with the size of short control messages in multicore processors. Out of those 80 bits, 20 are for the preamble (address and first bits of data) and 60 bits for the rest of the data. Without loss of generality, we assume a wireless datarate of 20 Gbps so that the preamble fits within a single clock cycle. The rest of data can be transmitted in $N=3$ more clock cycles after the ACK. It must be noted that this datarate requirement is well within the limits of achievable TR performance ($\sim$30\,Gbps) with minimal CCI that has been shown in prior work~\cite{bandara2023exploration, rodriguez2023}.

\begin{figure}[t]
\centering
\includegraphics[width=0.6\columnwidth]{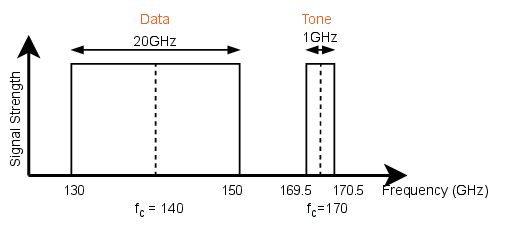}
%\vspace{-0.3cm}
\caption{Channel allocation strategy in TRMAC: data at 140\,GHz; tone signaling at 170\,GHz.}
\label{fig:channelallocation}
\end{figure}

We modified Multi2sim to integrate the TRMAC following the timing considerations above, and also taking into account that the tone signal prevents a node from starting the Phase I of a transmission during the Phase III of other nodes' transmissions. To sustain both the data and tone signaling requirements, we use the 140 GHz band with 20 GHz bandwidth for data transmission and the 170 GHz band for the tone signal, as shown in Figure~\ref{fig:channelallocation}. We further note that the addition of the tone signal leads to a very small energy overhead. According to prior work, a tone signal at the chip level can add approximately 2 mW of power consumption. This results in a 0.075 pJ/bit of efficiency considering the three nanoseconds during which it must be active for each 80 bits of data transfer, which is negligible compared to the 1-pJ/bit target efficiency of WNoC.

% Evaluation
\section{Evaluation}~\label{sec:results}

We evaluate the performance of the proposed TRMAC protocol in comparison with two baseline MAC strategies: Reliable-Broadcast-Sensing MAC protocol for WNoC (BRS) and Token Passing \cite{mestres2016mac, franques2021fuzzy}, under various synthetic traffic models and system configurations. All simulations were conducted using Multi2Sim~\cite{ubal2012multi2sim}, where TRMAC was implemented on a WNoC supporting spatial parallelism.

The protocols are compared in terms of latency and throughput as a function of the number of parallel transmissions, available antennas, cores, and traffic variations. TRMAC operates with spatially multiplexed links on a single frequency channel, while the baselines employ multi-channel configurations: BRS is a preamble-based MAC adapted for quasi-static wireless channels, and Token Passing creates a virtual ring among nodes using dedicated channels. To ensure fairness, we assign $N$ frequency channels to the baselines and evaluate TRMAC using $N$ simultaneous spatial transmissions on a shared frequency band. Simulation parameters are given in Table~\ref{tab:simpara}.

\begin{table*}[t]
\centering
\caption{Simulation parameters.}
\label{tab:simpara}
\resizebox{0.95\textwidth}{!}{%
\begin{tabular}{l l}
\toprule
\textbf{Parameter} & \textbf{Value} \\
\midrule
Number of antennas & 64--256 (one antenna per core) \\
Number of parallel transmissions (NPT) & 2--4 \\
Number of frequency channels & 1--4 \\
Network configuration & 80 bits per packet, 20-bit preamble \\
Synthetic traffic model & Hurst exponent $H$ = 0.5--1 \\
                        & Spatial hotspotness $\sigma$ = 0.1--10 \\
MAC protocols evaluated & TRMAC, BRS, Token Passing \\
Center frequency & 140 GHz (data), 170 GHz (tone) \\
Bandwidth & 20 GHz (data), 1 GHz (tone) \\
Data rate & 20 Gbps (TR with ASK modulation) \\
\bottomrule
\end{tabular}%
}
\end{table*}

\begin{figure*}[t]
\centering
\begin{subfigure}[t]{0.48\columnwidth}
\includegraphics[width=1\columnwidth]{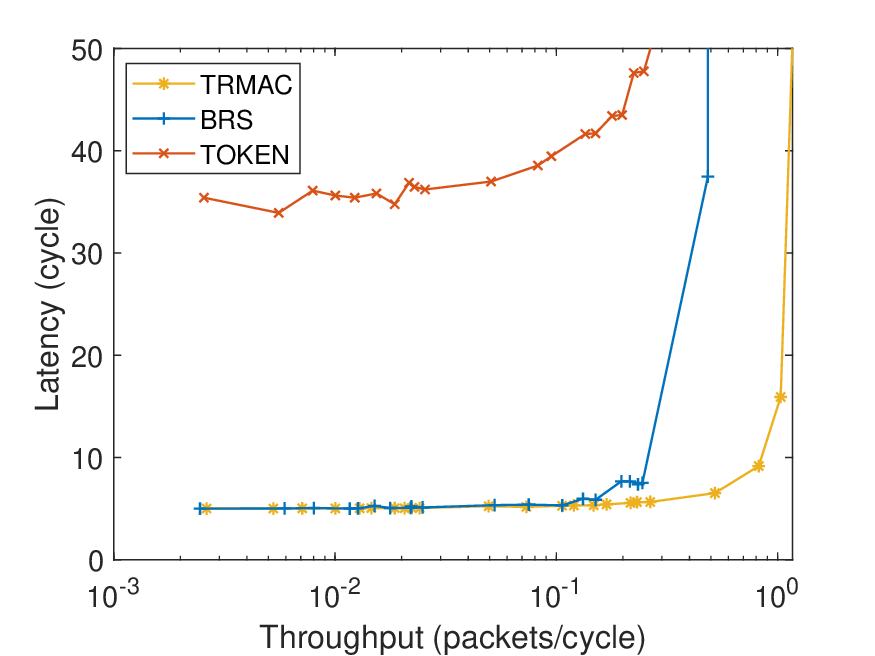}
\caption{}
\label{fig:singch}
\end{subfigure}
\begin{subfigure}[t]{0.48\columnwidth}
\includegraphics[width=1\columnwidth]{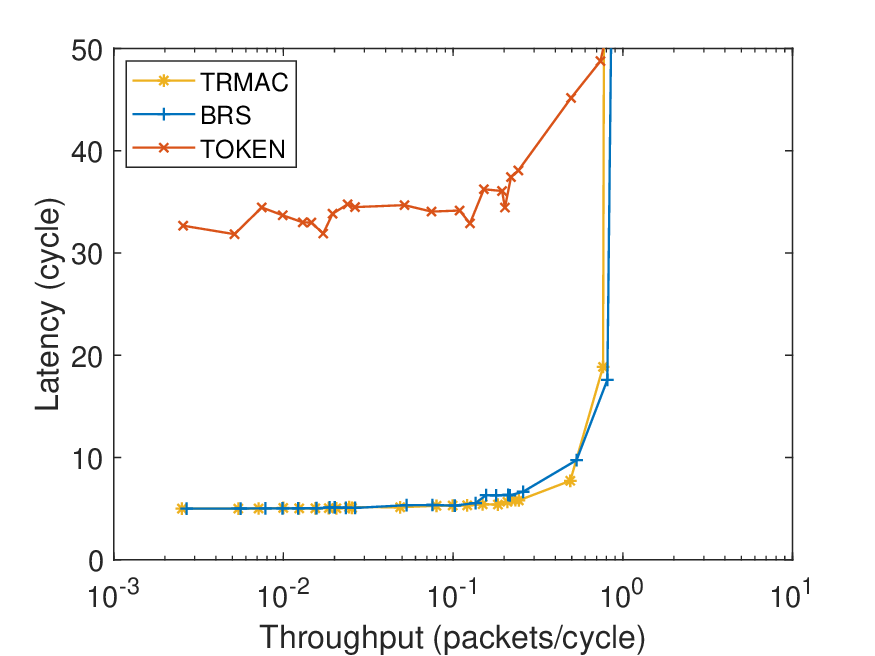}
\caption{}
\label{fig:mulnpt1}
\end{subfigure}
\begin{subfigure}[t]{0.48\columnwidth}
\includegraphics[width=1\columnwidth]{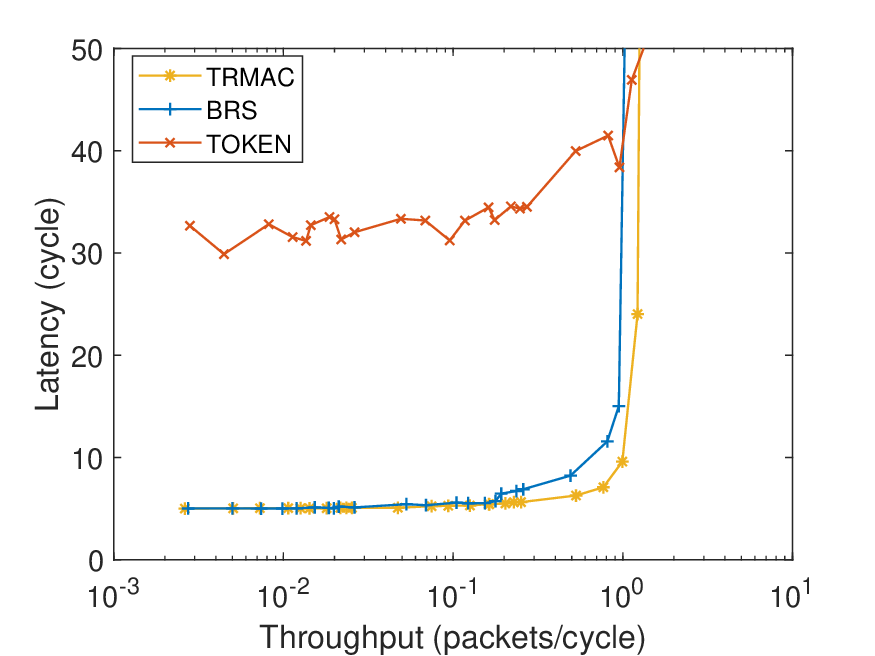}
\caption{}
\label{fig:mulnpt2}
\end{subfigure}
\begin{subfigure}[t]{0.48\columnwidth}
\includegraphics[width=1\columnwidth]{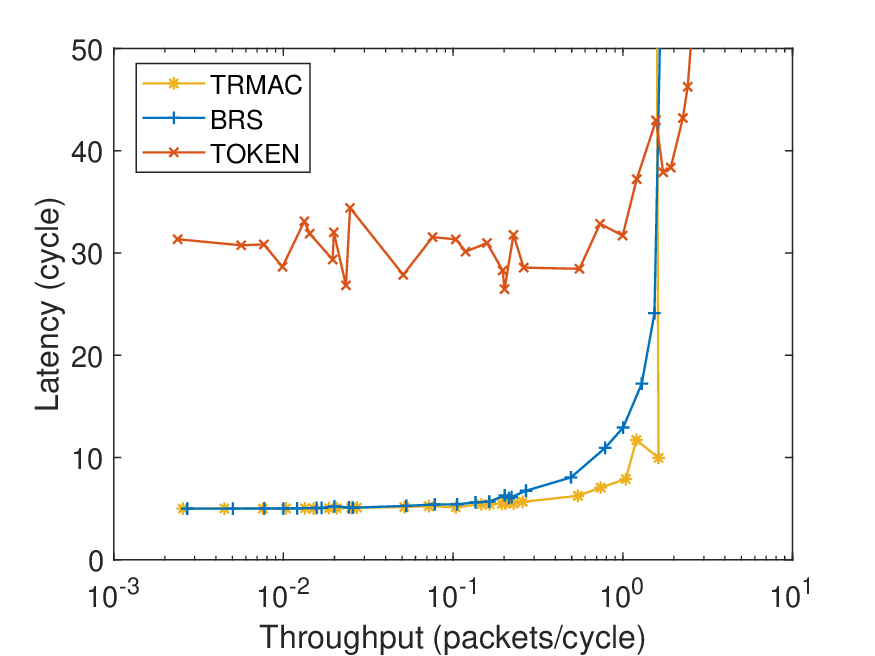}
\caption{}
\label{fig:mulnpt3}
\end{subfigure}
\caption{Latency-throughput curves for TRMAC and baselines with 64 cores using (a)  one channel for BRS and TOKEN / NPT=3 for TRMAC  (b) two channels for BRS and TOKEN / NPT=2 for TRMAC, (c) three channels for BRS and TOKEN  / NPT=3 for TRMAC (d) four channels for BRS and TOKEN / NPT=4 for TRMAC.}
\label{fig:npt}
\end{figure*}

\subsection{Number of Parallel Transmissions}

This subsection compares the performance of TRMAC against BRS and Token Passing, with respect to the number of parallel transmissions. The core premise of TRMAC is its ability to support multiple parallel transmissions (NPT) within a single frequency channel. To evaluate its effectiveness, we assess latency and throughput for NPT values from 2 to 4, under a synthetic traffic model with burstiness and hotspotting ($H=1$, $\sigma=0.5$).

We first observe the performance of TRMAC:3 NPT, with respect to single channel token/BRS. It is shown that in \mbox{Figure~\ref{fig:singch}}, with similar channel conditions the latency of TRMAC is low and throughput is high with a high performance gain in compared to the baselines. 

Consequently, we evaluate single-band TRMAC with multi-band token and BRS. \mbox{Figure~\ref{fig:mulnpt1}, Figure~\ref{fig:mulnpt2} and Figure~\ref{fig:mulnpt3}} show that TRMAC achieves comparable latency to BRS, and much better latency than Token passing, despite operating on a single frequency channel. In terms of throughput, TRMAC achieves saturation throughput equivalent to token passing and equal performance as BRS with less resource allocation. 
As the number of transmissions increases, latency in both TRMAC and BRS grows similarly due to increased contention. Token Passing exhibits the highest latency, which scales significantly with the ring length and added contention per token cycle.

These results confirm that TRMAC enables high throughput without the need for orthogonal frequency channels. Instead of reducing latency directly, it improves aggregate bandwidth by enabling concurrent transmissions in time and frequency. Moreover, TRMAC achieves performance parity with multi-channel protocols while avoiding the hardware overhead of multiple antennas and multi-band transceivers.

\begin{figure*}[t]
\centering
\begin{subfigure}[t]{0.48\textwidth}
\includegraphics[width=\textwidth]{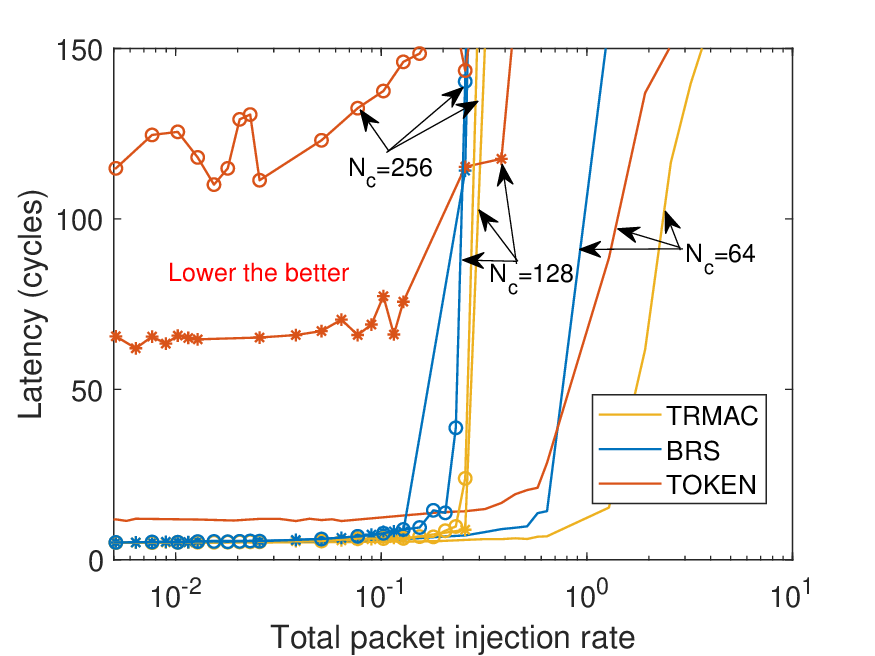}
\caption{Latency vs. packet injection rate.}
\label{fig:latency}
\end{subfigure}
\begin{subfigure}[t]{0.48\textwidth}
\includegraphics[width=\textwidth]{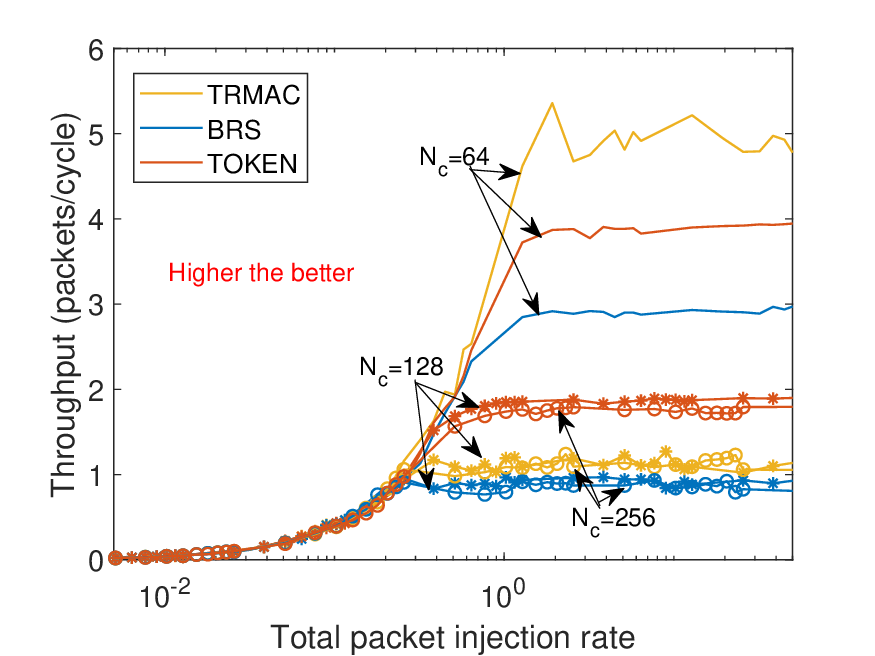}
\caption{Throughput vs. packet injection rate.}
\label{fig:throughput}
\end{subfigure}
\caption{TRMAC vs. BRS and Token Passing with increasing number of cores (64--256) using NPT=4/4 channels.}
\label{fig:numcores}
\end{figure*}

\subsection{Number of Cores}

We now evaluate the effect of increasing the number of cores (from 64 to 256) on protocol performance. We configure BRS and Token Passing with 4 frequency channels, while TRMAC uses NPT=4 on a single channel.

Figure~\ref{fig:numcores} shows that as the number of cores increases, total traffic increases proportionally. This leads to more collisions and an expansion of the token ring, degrading latency in Token Passing. Both TRMAC and BRS show lower latency at low injection rates, but Token Passing maintains higher throughput at high loads due to better collision avoidance. 
 
However, in BRS and TRMAC, throughput slightly decreases with more cores due to increased contention. Despite this, TRMAC performs better than BRS at high injection rates with a single frequency channel while BRS, TOKEN uses 4 channels, offering a favorable trade-off between latency and throughput as system size scales. 

\begin{figure}[t]
\centering
\includegraphics[width=0.45\textwidth]{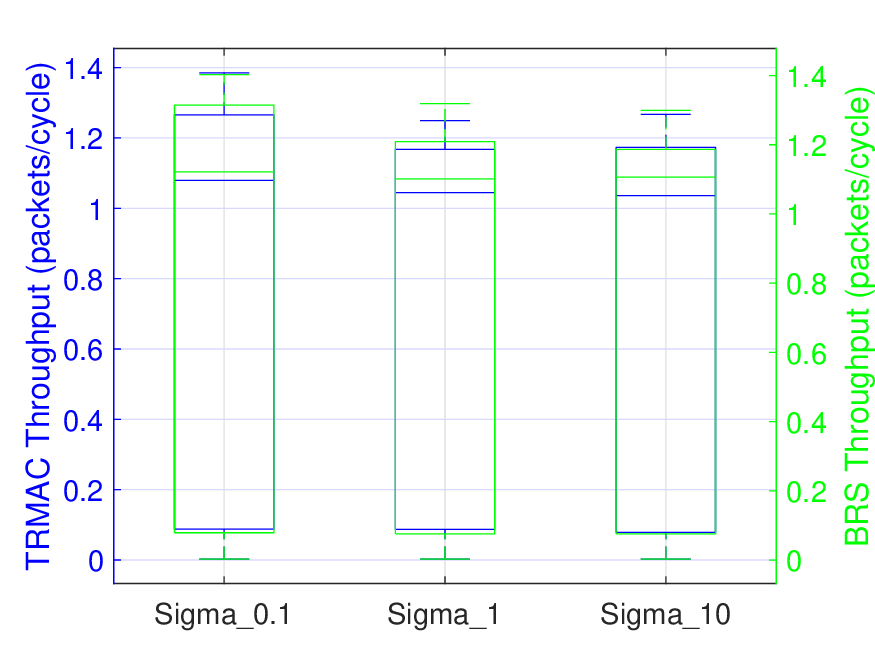}
%\label{fig:sigma1}
\includegraphics[width=0.45\textwidth]{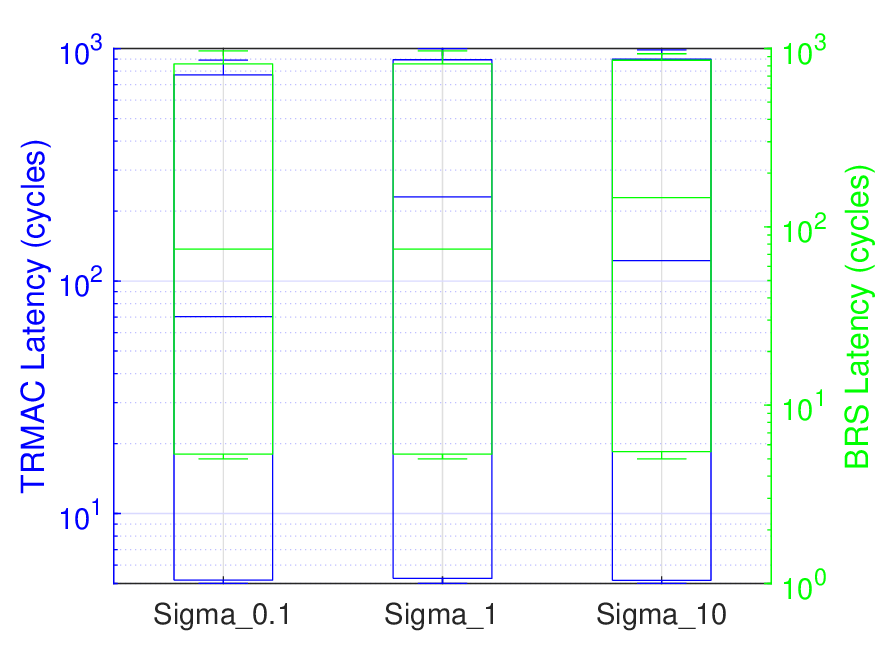}
%\label{fig:sigma2}
\includegraphics[width=0.45\textwidth]{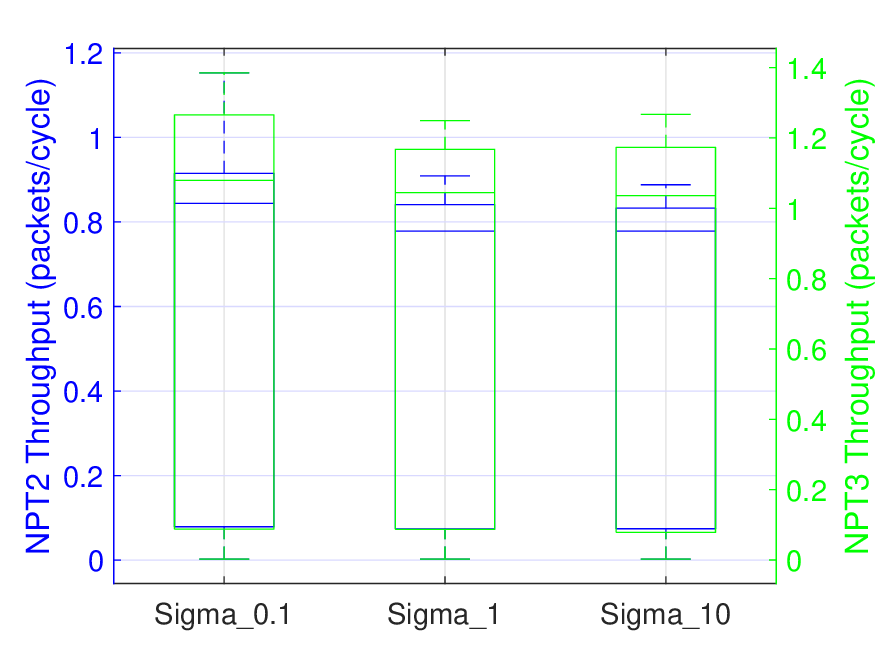}
%\label{fig:sigma3}
\includegraphics[width=0.45\textwidth]{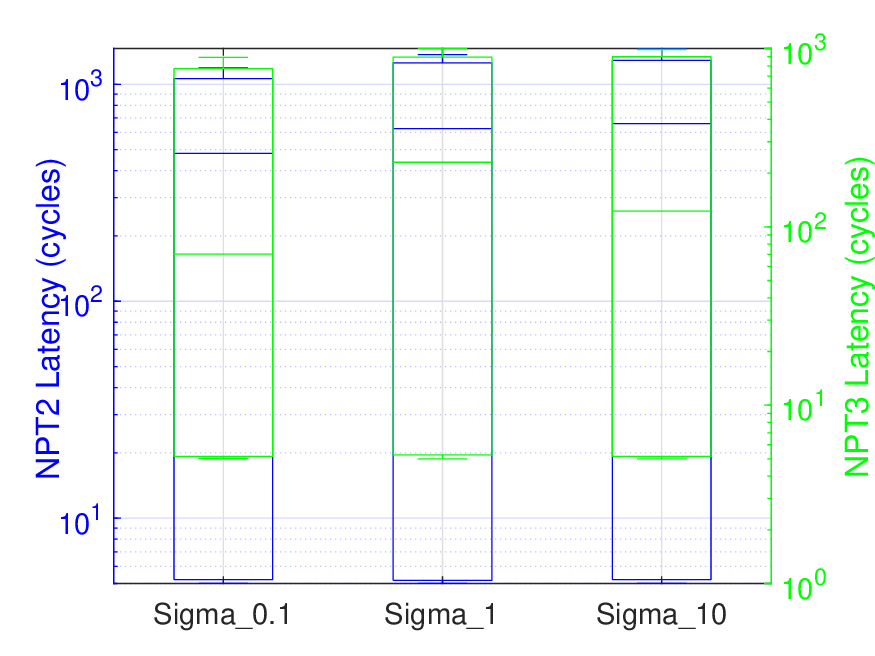}
%\label{fig:sigma4}
\vspace{-0.3cm}
\caption{Impact of spatial traffic distribution ($\sigma$) on the throughput (left) and latency (right) of TRMAC. Top plots compare the performance of TRMAC with one frequency channel and NPT=3 to that of BRS with three frequency channels. Bottom plots compare the performance of TRMAC with NPT=2 to that of TRMAC with NPT=3.}
\label{fig:spatial1}
\end{figure}

\begin{figure}[!t]
\centering
\begin{subfigure}[t]{0.45\textwidth}
\includegraphics[width=1\textwidth]{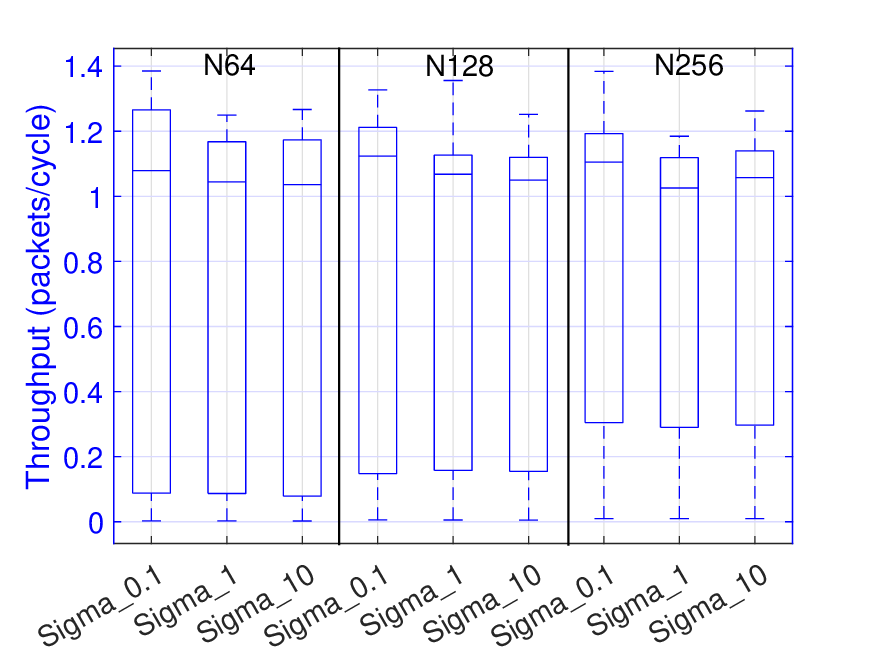}
\label{fig:sigma5}
\end{subfigure}
\begin{subfigure}[t]{0.45\textwidth}
\includegraphics[width=1\textwidth]{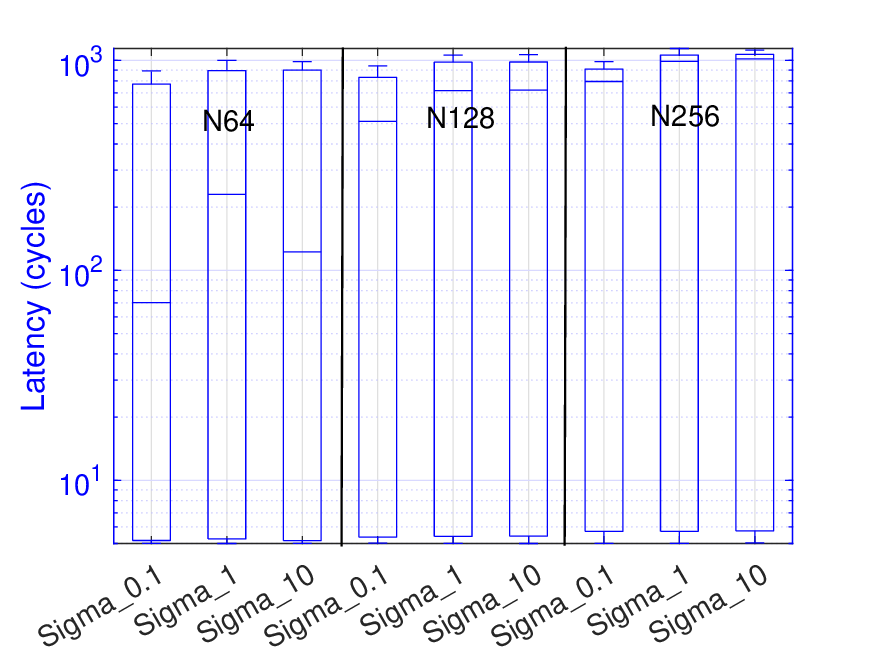}
\label{fig:sigma6}
\end{subfigure}
\vspace{-0.3cm}
\caption{Impact of spatial traffic distribution ($\sigma$) on the throughput (left) and latency (right) of TRMAC when increasing number of cores with one frequency channel and NPT=3.}
\label{fig:spatial2}
\end{figure}

\subsection{Traffic Sensitivity Analysis}
We examine the sensitivity of TRMAC to spatial and temporal traffic variations using the synthetic traffic model. We analyze the standard deviation $\sigma$ (representing spatial distribution) and the Hurst exponent $H$ (representing temporal burstiness) across broad ranges: $\sigma=0.1$ - 10 and $H=0.5$ - 1. Results are compared with BRS, as both are designed for on-chip environments and exploit preamble-based access. The box plot representations follow a data set with the packet injection rates from 0.00001-1. The box itself represents the middle 50\% data from first to the third quartile, while the line inside the box shows the middle value of the data set. The outside whiskers show the largest and smallest values within the range of 1.5 times the inter-quartile range.

\subsubsection{Spatial Distribution}

The spatial distribution determines the traffic load imbalance among nodes. Lower $\sigma$ corresponds to more localized hotspot traffic, while higher values imply uniform traffic across nodes.

The top plots of Figure~\ref{fig:spatial1} compares TRMAC with BRS across the full $\sigma$ range. Both protocols perform similarly under varying distribution, confirming TRMAC’s ability to handle both hotspot and evenly distributed traffic with efficient spectrum allocation such that having single frequency channel and NPT=2, 3 multiple parallel transmissions.

In Figure~\ref{fig:spatial1}, TRMAC demonstrates increased throughput (bottom left) and reduced latency (bottom right)  as NPT increases. More spatial channels allow better traffic dispersion. However, at higher $\sigma$, throughput slightly drops due to more concurrent access attempts causing contention.

Figure~\ref{fig:spatial2} shows that TRMAC retains stable performance as the number of cores increases from 64 to 256 under varying $\sigma$. Throughput and latency trends remain consistent, confirming robustness under spatially imbalanced traffic.

\subsubsection{Temporal Burstiness}
The Hurst exponent $H$ represents temporal correlations in traffic. Higher values indicate more bursty behavior, while lower values suggest uniform injection.

\begin{figure}[t]
\centering
\includegraphics[width=0.45\textwidth]{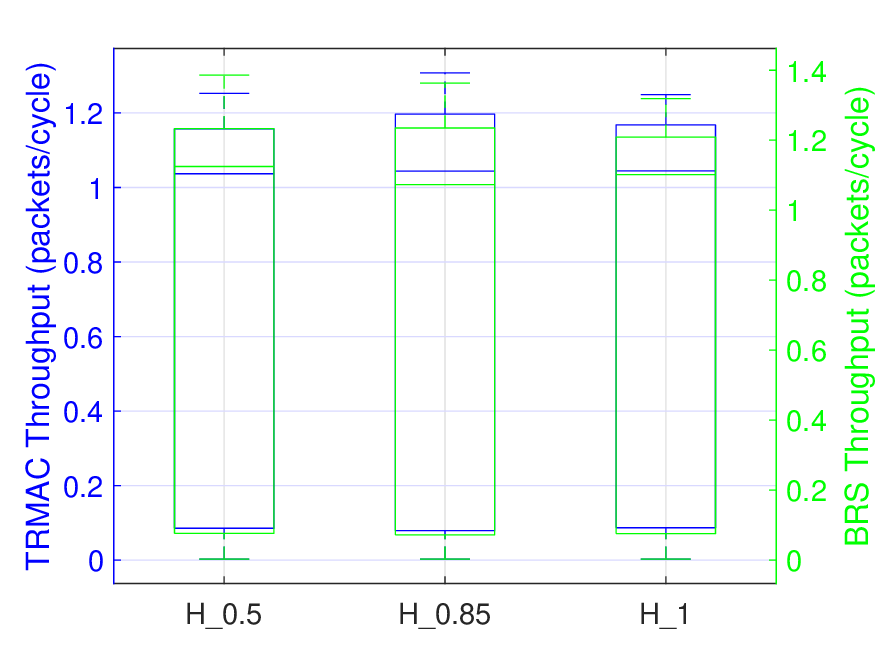}
\label{fig:H1}
\includegraphics[width=0.45\textwidth]{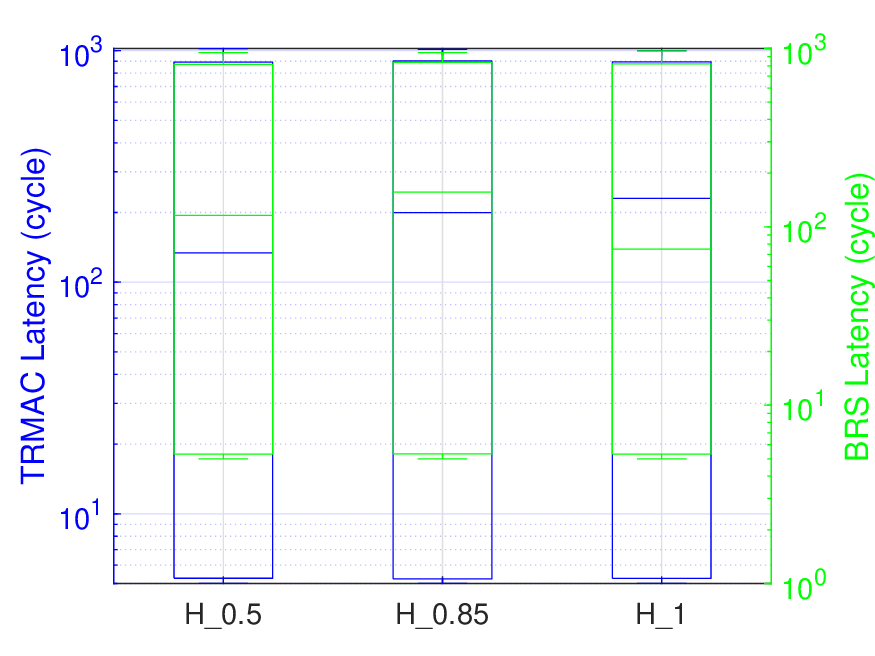}
\label{fig:H2}
\includegraphics[width=0.45\textwidth]{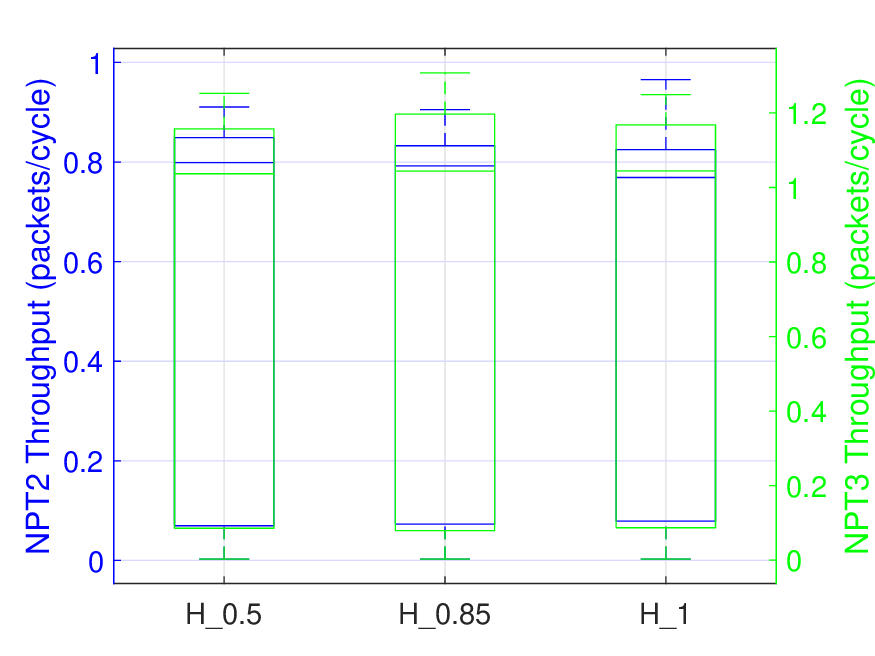}
\label{fig:H3}
\includegraphics[width=0.45\textwidth]{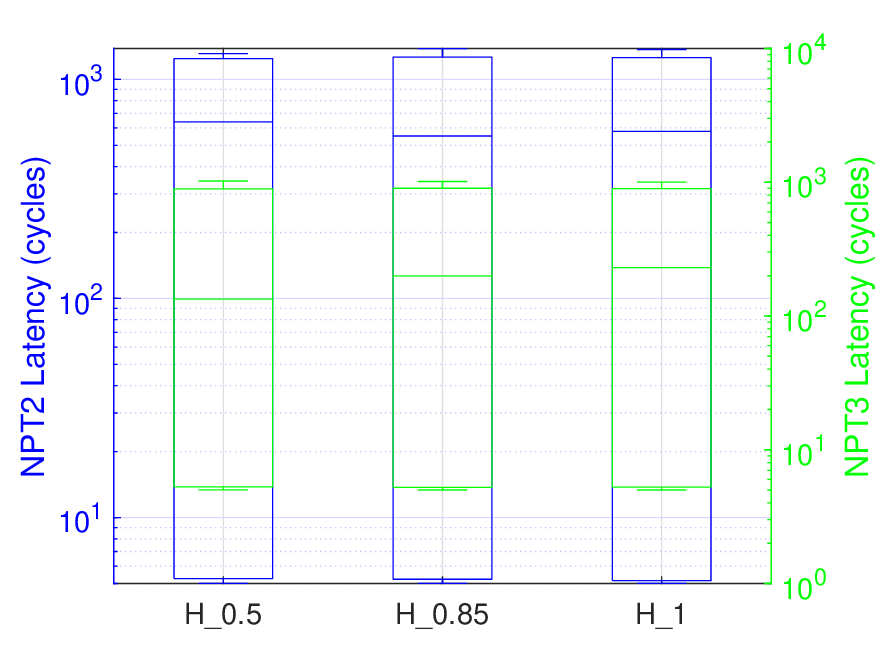}
\label{fig:H4}
\vspace{-0.3cm}
\caption{Impact of spatial traffic burstiness ($H$) on the throughput (left) and latency (right) of TRMAC. Top plots compare the performance of TRMAC with one frequency channel and NPT=3 to that of BRS with three frequency channels. Bottom plots compare the performance of TRMAC with NPT=2 to that of TRMAC with NPT=3.}
\label{fig:temporal1}
\end{figure}

Figure~\ref{fig:temporal1} (top plots) compares TRMAC with BRS under bursty traffic. Both protocols show nearly identical throughput trends, while the latency (top right) of TRMAC is smaller than BRS. This performance  validates robustness of TRMAC against temporal injection variability and proving that it performs similarly with spatial diversity with one channel by using 1/2, 1/3 of resources compared to baseline protocols.

\begin{figure}[!t]
\centering
\begin{subfigure}[t]{0.45\textwidth}
\includegraphics[width=1\textwidth]{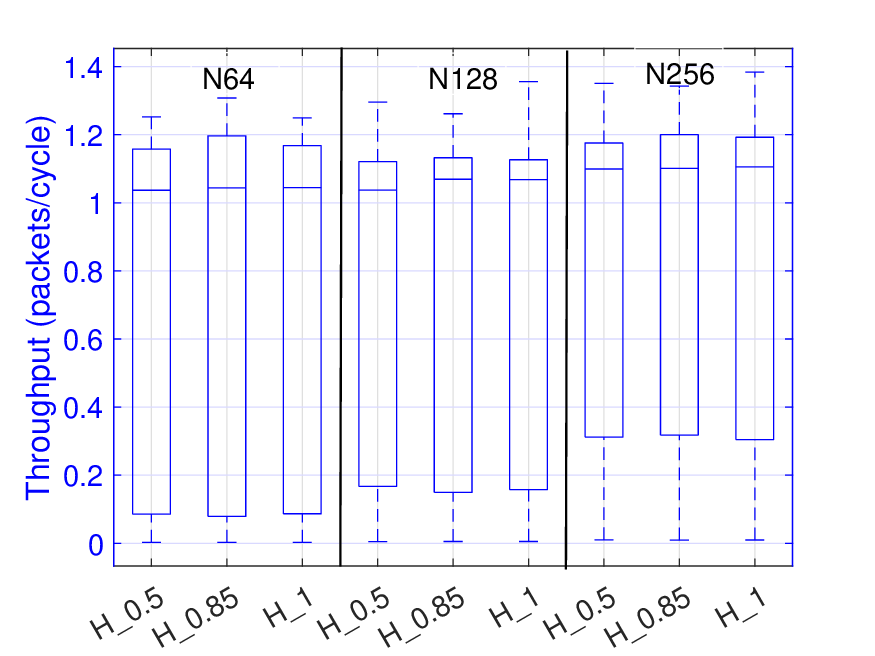}
\label{fig:H5}
\end{subfigure}
\begin{subfigure}[t]{0.45\textwidth}
\includegraphics[width=1\textwidth]{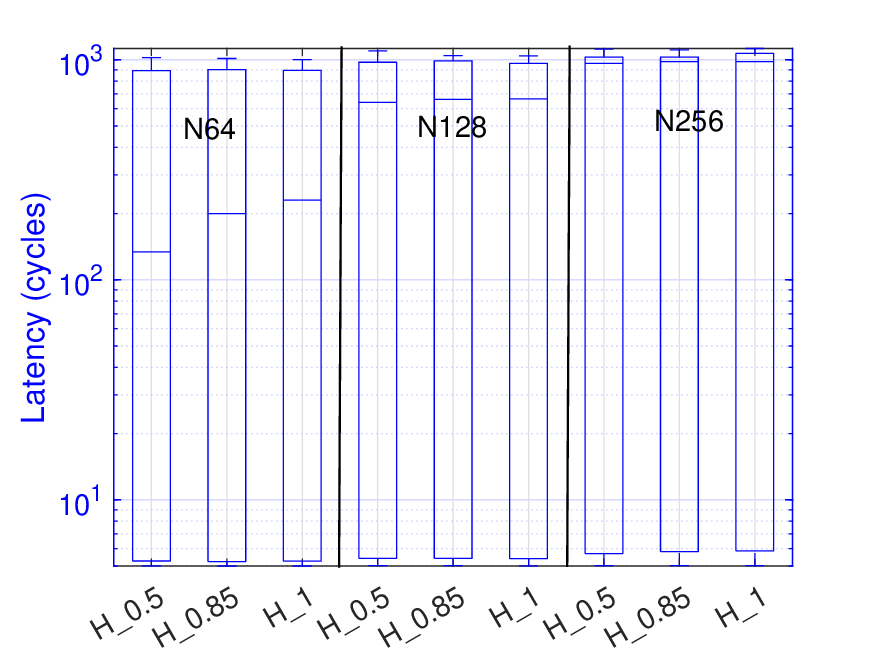}
\label{fig:H5}
\end{subfigure}
\vspace{-0.3cm}
\caption{Impact of traffic burstiness ($H$) on the throughput (left) and latency (right) of TRMAC when increasing number of cores with one frequency channel and NPT=3.}
\label{fig:temporal2}
\end{figure}

In Figure~\ref{fig:temporal1} (bottom plots), TRMAC maintains similar throughput and latency trend for NPT=2 and NPT=3, regardless of $H$. Figure~\ref{fig:temporal2} shows that even as core count increases, the performance of TRMAC remains stable across different burstiness levels.

% Related Work
\section{Related Work}~\label{sec:relatedwork}

In this section, we review existing WNoC protocols, focusing on their medium access techniques, scalability, architectural design, and adaptability to varying traffic patterns. \mbox{Table \ref{tab:relatedwork}} summarizes several closely related and recent MAC protocols for WNoC, offering a comparative perspective on their performance characteristics.

\subsection{Token Passing and CSMA}
Several MAC protocols based on token-controlled data flow have been explored in prior work~\cite{ chang2012, mansoor2016}, where the node holding the token gains exclusive access to the communication medium during designated time slots, eliminating the need for global synchronization. While token passing schemes typically offer high throughput under controlled conditions, they struggle to scale under hotspot or highly dynamic traffic patterns.

The fuzzy token MAC protocol proposed in~\cite{franques2021fuzzy} presents an adaptive variant that dynamically adjusts transmission probabilities based on traffic intensity, offering low-latency, collision-free transmissions under high load conditions. It incorporates a contention-aware mechanism that adapts the token region based on local collision statistics and load, thereby balancing performance under variable traffic.

\begin{table*}[t!]
\begin{center}
\caption{Summary of the design traits of existing protocols}
\label{tab:relatedwork}
\renewcommand{\arraystretch}{2}
\resizebox{\columnwidth}{!}
{\begin{tabular}{|m{2.5cm}|m{2.5cm}|m{2.5cm}|m{2.5cm}|m{2.5cm}|m{7.5cm}|}
\hline 
\textbf{Work} & \textbf{Architecture} & \textbf{Multiplexing} & \textbf{Scalability} & \textbf{Traffic Awareness} & \textbf{Remarks} \\ [1.5ex]
\hline
Shamim et al.~\cite{Shamim2017} & Dynamic & N/A & High & Low& Energy efficient wireless links with token passing.\\
\hline
Mansoor et al.~\cite{mansoor2015reconfigurable} & Dynamic & TDMA & Medium & High& Time-slot allocation based on traffic prediction.  \\
\hline
Franques et al.~\cite{franques2021fuzzy} & Dynamic & N/A & High & High& Dynamic traffic control with token passing.\\
\hline
Mestres et al.~\cite{mestres2016mac} & Dynamic & N/A & High & Low& Simple and highly adaptable for on-chip/package scenario.\\
\hline
Vijayakumaran et al.~\cite{Vijayakumaran2014} & Static & CDMA & High & High& Analytical evaluation of data transfer for on chip networks with CDMA, which leads to complex transceiver design.\\
\hline
Matolak et al.~\cite{MatolakF2012} & Static & FDMA, TDMA & High & Medium & Hybrid wired-wireless architecture with low power and increased area efficiency. \\
\hline
DiTomaso et al.~\cite{DiTomaso2015} & Static & TDMA & High & High& Adaptable hybrid wired-wireless architecture improving energy efficiency\\
\hline
Jog et al.~\cite{jog2021one} & Dynamic & TDMA & High & High& Deep reinforcement learning (DRL) based dynamic traffic control and a DRL framework to operate within precise timing, area and power constraints for on-chip communication networks. \\
\hline
Ollé et al.~\cite{Ollé2023} & Dynamic & FDMA & High & High& Exploring multi-channel medium access with existing protocols with dynamic traffic, random access and scalability. \\
\hline
Rout et al.~\cite{Rout2023}& Dynamic& N/A & High & High& Congestion aware MAC with dynamic channel allocation according to the traffic requirements. \\
\hline
\hline
\textbf{TRMAC (this work)} & \textbf{Dynamic} & \textbf{Spatial} & \textbf{High} & \textbf{High}& Time reversed parallel transmissions with spatial diversity in same frequency and time. \\
\hline
\end{tabular}%
}
\end{center}
\end{table*}

A predictive time-slot allocation MAC protocol is introduced in~\cite{mansoor2016}, enabling on-demand medium access based on the anticipated bandwidth requirements. On the other hand, CSMA-based wireless NoC designs have been proposed in~\cite{dai2015}, while~\cite{mestres2016mac} presents a reliable CSMA-inspired MAC with Negative Acknowledgment (NACK) for collision detection, offering a middle ground between CSMA/CA and CSMA/CD.

Recent work in~\cite{Ollé2023} compares the performance of multi-channel MAC protocols, including token passing and the broadcast-aware CSMA design from~\cite{mestres2016mac}, under both statistical and random traffic injection patterns across varying node counts.

Although these protocols address latency and contention to varying degrees, their reliance on complex coordination or orthogonal channel usage limits scalability in massive multi-core or chiplet-based systems. In such contexts, lightweight MAC solutions capable of supporting concurrent transmissions on shared channels, such as TRMAC, offer greater scalability and efficiency.

\subsection{Multiplexing}

Multiplexing strategies have long been a foundational approach in WNoC MAC designs. Protocols based on time, frequency, or code multiplexing (TDMA, FDMA, and CDMA) have been proposed to ensure low-contention data transmission through orthogonal resource allocation.

A scalable hybrid TDMA-FDMA architecture is presented in~\cite{Ganguly2011}, with the potential to integrate emerging technologies such as carbon nanotube antennas. In~\cite{DiTomaso2015}, a time-division-based wired-wireless system is evaluated, demonstrating the feasibility of dynamic transceiver adaptation to achieve high data rates and low energy consumption.

CDMA-based approaches are explored in~\cite{Vijayakumaran2014} and~\cite{duraisamy2015enhancing}, offering the benefits of parallelism through code orthogonality. However, the requirement for complex transceiver designs capable of handling orthogonal spreading codes limits the practicality of CDMA in dense, high-speed environments.

As core and chiplet counts increase, the hardware complexity of strict multiplexing schemes becomes a critical bottleneck, motivating the need for alternative techniques like spatial multiplexing via TR filters, as employed in TRMAC.

\subsection{Architecture}

Hybrid wired-wireless NoC architectures have been extensively explored to mitigate the limitations of long-range wired interconnects by integrating low-latency wireless links. These designs often toggle between MAC mechanisms such as random access and token passing~\cite{mansoor2015reconfigurable} or target application-specific goals such as shared-memory consistency and fine-grained synchronization in multi-core systems~\cite{franques2021widir}.

However, many of these protocols are either tied to specific architectural assumptions, tailored to particular application domains, or require intricate control logic to adapt to varying conditions. In contrast, TRMAC offers a generalized, adaptable MAC framework that leverages the quasi-static and reverberant properties of the on-chip wireless channel. By exploiting spatial reuse and energy focusing via TR filters, TRMAC can support scalable, low-complexity MAC operations across a wide range of system sizes and traffic conditions, making it a promising candidate for future many-core and multi-chiplet environments.

% Conclusion
\section{Discussion and Limitations}
\label{sec:discussion}
TRMAC is a novel protocol which is designed for on-chip wireless networks that exploits TR filters for mitigating the channel impairments and to utilize parallel transmissions both in time and space. Despite of the spatial diversity, as TR filters do not prevent the transmission collisions, TRMAC fills the gap of coordination and control of TR precoded data transmissions. Unlike conventional MAC protocols that rely on frequency division or centralized arbitration, TRMAC operates using a single frequency channel while supporting multiple concurrent transmissions, thereby reducing both hardware complexity and spectral overhead. 

The protocol prevents the collisions and addresses the deafness problem by transmitter-side channel sensing, acknowledgment feedback and tone-signaling in each phase of the proposed solution, by considering all the possible collision events that could occur with multiple concurrent transmitters and receivers operating in same time and frequency. Through comprehensive set of simulations it is shown that TRMAC maintains high throughput and low latency with multiple spatial channels, scales efficiently with increasing number of cores, adapt robustly to spatial and temporal traffic variations, while performing similar to or better than baseline protocols such as BRS and token passing with efficient resource allocation.

However, channel symmetry plays an important role in TRMAC, where the CIR of each link is composed with similar characteristics. Even though it is unlikely to obtain perfect symmetric channels with equal amplitude and phase, there is a possibility where the CCI could be increased due to spatial similarities. Yet, with the proposed TRMAC protocol, we pre-characterize each CIR prior to the transmission and set energy thresholds and NPT based on the evaluated CIRs. Channel symmetry could potentially limit the NPT, and the trade-off will be in between the expected NPT and the BER of the collective communications. 

In terms of implementation, TR filters have to resemble a precise time-reversed CIR of the characterised channel response to mimic the ideal filter and to achieve perfect energy focusing. However, with the limitation of achievable sampling rates (up to a few tens of GSps), imperfect TR filters may negatively impact on the performance of the system. Additional limitations may come from the amount of memory required to store the filters. In extremely reverberant environments, the length of the channel could be several tens of nanoseconds long. Fortunately, even assuming very high sampling frequencies, the total duration of the filter could be covered with less than a thousand samples. In this context, considering a large filter with 1000 samples and high numerical precision (16 bits), around 16 filters can be stored in a 32-KB L1 cache memory beside each antenna. Considering modern memory designs, such cache memories have a very low access latency and would therefore not impact the performance of the protocol. For scaling to more filters per antenna, one could reduce the number of samples or numerical precision at the expense of a lower TR quality. 

Moreover, TR filter calibration plays a vital role in error correction due to mismatches occurring in the channel response. Indeed, the real-time CIR could be deviated from the design phase due to discrepancies occurring due to chip manufacturing and packaging. Hence, feedback-based mechanisms can be employed for filter calibration to adjust the initial measurements, which is a common practice in testing the chips \mbox{\cite{trcommag}}.

\section{Conclusion}~\label{sec:conclusion}

This work introduced TRMAC, a cross-layer MAC protocol designed for wireless communication in dense multi-core and chiplet-based systems. By leveraging the spatial focusing properties of TR in quasi-deterministic on-chip wireless channels, TRMAC enables multiple parallel transmissions on a shared frequency band without complex coordination or orthogonal resource allocation. The protocol uses pre-characterized channel impulse responses and energy thresholds to manage contention with low overhead. Physical-layer simulations and system-level evaluations show that TRMAC matches or exceeds the performance of existing MAC protocols such as BRS and token passing, achieving similar or better performance with 1/2, 1/3 of the spectrum
%achieving high throughput, low latency, and efficient spectrum reuse, 
even under bursty and hotspot traffic patterns. TRMAC share the spectral resources by multiplying the throughput with low latency. Its ability to scale with core count while avoiding the hardware complexity of multi-band transceivers makes it well-suited for future many-core and chiplet-integrated platforms. TRMAC represents a significant step toward exploiting spatial multiplexing in on-chip wireless networks and offers a robust, low-complexity solution for next-generation processor interconnects. Future directions include integrating TRMAC into heterogeneous systems, supporting QoS-aware traffic control, and aligning with chiplet interconnect standards for modular computing.
% References
\bibliographystyle{elsarticle-num} 
\bibliography{references}

\end{document}